\documentclass[12pt,preprint]{aastex}
\usepackage{natbib}
\usepackage{psfig}
\usepackage{epsfig}

\shorttitle{On the stability of thick accretion disks around black holes}
\shortauthors{Font \& Daigne}

\begin{document}

\title{On the stability of thick accretion disks around black holes}

\author{Jos\'e A. Font}
\affil{Departamento de Astronom\'{\i}a y Astrof\'{\i}sica, 
Universidad de Valencia \\ Edificio de Investigaci\'on, Dr. Moliner 50, 
46100 Burjassot (Valencia), Spain}

\and

\author{Fr\'ed\'eric Daigne}
\affil{CEA/DSM/DAPNIA, Service d'Astrophysique, C.E. Saclay, 91191 Gif sur 
Yvette Cedex, France\\
Present address: Institut d'Astrophysique de Paris, 98 bis boulevard Arago, 
75014 Paris France}

\begin{abstract}
Discerning the likelihood of the so-called runaway instability of thick 
accretion disks orbiting black holes is an important issue for most models 
of cosmic gamma-ray bursts. To this aim we investigate this phenomenon by 
means of time-dependent, hydrodynamical simulations of black hole plus torus 
systems in general relativity. The evolution of the central black hole is 
assumed to be that of a sequence of Kerr black holes of increasing mass 
and spin, whose growth rate is controlled by the transfer of mass and angular 
momentum from the material of the disk spiralling in through the event horizon 
of the black hole. The self-gravity of the disk is neglected. We find that 
when the black hole mass and spin are allowed to increase, constant angular 
momentum disks undergo a runaway instability on a dynamical timescale (a few 
orbital periods). However, our simulations show that a slight increase of the 
specific angular momentum of the disk outwards has a dramatic stabilizing effect. 
Our results, obtained in the framework of general relativity, are in broad 
agreement with earlier studies based both on stationary models and on 
time-dependent simulations with Newtonian and pseudo-Newtonian gravitational 
potentials.
\end{abstract}


\section{Introduction}

The so-called runaway instability of thick accretion disks orbiting black
holes was first noticed by \citet{abramowicz:83}. The origin of the instability is
a dynamical process by which the cusp of a disk  initially filling its Roche lobe
moves outwards due to mass transfer from the disk to the accreting black hole. This
process leads to the complete destruction of the disk on a dynamical timescale.
\citet{abramowicz:83} found that the runaway instability occurs for a large range
of parameters such as the disk-to-hole mass ratio, $M_{\mathrm{D}}/M_{\mathrm{BH}}$,
and the location of the disk inner radius. More detailed studies followed, most 
of which are based on stationary models in which a fraction of the mass and angular 
momentum of the initial disk is transferred to the black hole. The new gravitational 
field allows to compute the new position of the cusp, which controls the occurrence 
of the runaway instability.

The main conclusion of those studies is that the very existence of the instability 
remains uncertain. Taking into account the self-gravity of the disk seems to favor 
the instability, as shown from both, studies based on a pseudo-potential for the 
black hole \citep{khanna:92,masuda:98} and from fully relativistic calculations 
\citep{nishida:96a}. However, the rotation of the black hole has a stabilizing effect 
\citep{wilson:84,abramowicz:98}, and the same happens when the disks are built with 
non-constant distributions of the specific angular momentum (increasing outwards) 
\citep{daigne:97,abramowicz:98}.

In a recent paper \citep{font:02} we presented the first time-dependent, hydrodynamical
simulations in general relativity of the runaway instability of constant angular 
momentum thick disks around a Schwarzschild black hole. The self-gravity of the disk 
was neglected and the evolution of the central black hole was assumed to follow
a sequence of Schwarzschild black holes of increasing mass. We found that by 
allowing the mass of the black hole to grow the runaway instability appears on 
a dynamical timescale, in agreement with previous estimates from stationary 
models. For a black hole of $2.5\ \mathrm{M_{\odot}}$ and disk-to-hole mass 
ratios between 1 and 0.05, our simulations showed that the timescale of the 
instability never exceeds $1\ \mathrm{s}$ for a large range of mass fluxes and 
it is typically a few 10 ms. Similar results for different initial data have been 
recently reported by \citet{zanotti:02}.

In this {\it Letter} we extend our study to the most interesting
case of {\it non-constant} angular momentum disks. The main motivation of our work 
is to check through time-dependent simulations in general relativity whether such
distributions have indeed the stabilizing effect previously found in 
non self-gravitating stationary models \citep{daigne:97,abramowicz:98,lu:00}.

\begin{table*}
  \caption{Model parameters\label{table:params}}
\begin{center}
\vspace*{-2ex}

  \begin{tabular}{cccccccccc}
    \hline \hline
    Model & $ M_{\mathrm{BH}} $ & $ M_{\mathrm{D}}/M_{\mathrm{BH}} $ & $ \alpha $ &
            $ \Delta W_{\mathrm{in}}/c^{2} $ & $ \dot{m}_{\mathrm{stat}} $
          & $ r_{\mathrm{cusp}} $ & $ r_{\mathrm{center}} $ 
          & $ t_{\mathrm{orb}} $ & $ t_{\mathrm{run}}/t_{\mathrm{orb}} $ \\
          & [M$_{\odot}$] & & & & [M$_{\odot}$/s] & & & [ms] & \\
    \hline
    A & 2.5 & 1.0 & 0.    & $3.81 \times 10^{-2}$ & $\sim 25.$ & 4.896 & 7.592 & 1.62 & $\sim 4$ \\
    B & 2.5 & 1.0 & 0.1   & $2.26 \times 10^{-2}$ & $\sim 2.5$ & 5.225 & 9.982 & 2.44 & stable \\
    \hline
    C & 2.5 & 0.1 & 0.    & $2.60 \times 10^{-2}$ & $\sim 5.9$ & 4.962 & 7.459 & 1.58 & $\sim 8$\\
    D & 2.5 & 0.1 & 0.075 & $1.87 \times 10^{-2}$ & $\sim 1.3$ & 5.243 & 8.964 & 2.08 & stable \\
    \hline
  \end{tabular}
\end{center}
\vskip -0.5truecm
\end{table*}

\section{Numerical framework}

A given initial state of a thick disk orbiting a black hole is determined by five
parameters: the black hole mass $M_\mathrm{BH}$, the specific angular
momentum in the equatorial plane $l$, the potential barrier at the inner edge 
of the disk $\Delta W_\mathrm{in}=W_\mathrm{in}-W_\mathrm{cusp}$, the polytropic 
constant $\kappa$ and the adiabatic index $\gamma$ of the equation of state 
(EoS). We assume a specific law of rotation of the disk in which the angular 
momentum in the equatorial plane increases outwards as a positive power law 
$l\propto r^{\alpha}$. The specific values we have considered for the exponent
$\alpha$ are listed in Table 1. The initial equilibrium configurations are built 
in various steps (see \citet{daigne:02} for details): first, from the coefficients 
of the Schwarzschild metric and the radial profile of $l$ in the equatorial plane, 
the angular momentum distribution outside the equatorial plane is computed by 
solving the equations describing the constant $l$ surfaces, i.e. the von Zeipel 
cylinders. Then, the total specific energy $u_{t}$ is obtained from the distribution 
of $l$ via the normalization of the 4-velocity, $u_{\mu}u^{\mu}=-1$. The integral 
form of the relativistic Euler equation allows then to compute the quantity 
$W=-\int_{0}^{P}dP/w$, where $P$ is the pressure and $w$ the enthalpy. The 
Newtonian limit of $W$ is the total (gravitational plus centrifugal) potential. 
The density and the pressure can be easily computed from $W$ using the EoS. 
The geometrical structure of the equipotential surfaces is comparable to that of 
the Roche lobes of a binary system. In particular, there is a cusp where 
mass transfer from the disk to the black hole is possible. 

As we did in our previous investigation \citep{font:02} we focus on models which 
are expected to exist at the central engine of gamma-ray bursts (GRBs), formed 
either after the coalescence of a compact binary system (composed of two neutron
stars or a neutron star and a stellar-mass black hole), or after the gravitational 
collapse of a massive star (see e.g. \citet{woosley:01}). Taking into account the 
various results from numerical simulations of those systems
\citep{ruffert:99,kluzniak:98,macfadyen:99,shibata:00,aloy:00} we fix the mass of the
black hole to $M_\mathrm{BH}=2.5$ M$_{\odot}$. The angular momentum $l$ is adjusted 
to yield a disk-to-hole mass ratio of 1 (models A and B) or of 0.1 (models C and D). 
The latter case is more realistic, being the parameters closer to those 
inferred from simulations of compact binary mergers. Furthermore, neglecting the 
disk self-gravity is also much better justified for $M_\mathrm{D}/M_\mathrm{BH}=0.1$. 
An even more realistic model would assume an initially rotating black hole, as both 
in collapsars and mergers the hole is expected to form with an initial spin parameter
$\ga 0.5$. Such models will be considered in a forthcoming study.

The EoS adiabatic index and polytropic constant are, respectively, $\gamma=4/3$ 
and $\kappa=4.76\times 10^{14}\ \mathrm{cgs}$. This corresponds to an EoS dominated 
by the contribution of relativistic degenerate electrons (the typical density 
in the disk is $\sim 10^{11}$-$10^{12}\ \mathrm{g/cm^{3}}$). The gravitational 
potential at the inner edge of the disk is fixed to $\Delta W_{\mathrm{in}}=0.75 
|W_{\mathrm{cusp}}|$ in models A and B, and $0.5|W_{\mathrm{cusp}}|$ in models C and 
D, which corresponds to a finite disk ($W_\mathrm{in}<0$), initially overflowing 
its Roche lobe ($\Delta W_\mathrm{in} > 0$) so that mass transfer is possible at 
the cusp. The parameters describing the models are summarized in Table 1.

The equilibrium initial data are evolved in time using the same relativistic 
hydrodynamics code employed in \citet{font:02}. This code integrates the 
non-linear hyperbolic system of the general relativistic hydrodynamic equations 
in a Kerr spacetime, using Boyer-Lindquist $(t,r,\theta,\phi)$ coordinates and 
under the restriction of axisymmetry, $\partial_{\phi}=0$. The numerical scheme, 
based on an approximate Riemann solver, is second-order accurate in both space and 
time due to the use of a piecewise linear reconstruction algorithm and a 
conservative, two-step Runge-Kutta scheme for the time update. The 
simulations use a numerical grid of $400\times 100$ zones in $r$ and $\theta$,
respectively. The radial grid is logarithmically spaced to account for 
sufficient resolution near the horizon ($r_{\mathrm{min}}=2.12M_{\rm{BH}}$, 
$\Delta r_{\mathrm{min}}=1.99\times 10^{-2}M_{\rm{BH}}$; $G=c=1$) and includes 
the whole disk. 

The procedure we follow to account for the spacetime dynamics is equivalent 
to the one employed in \citet{font:02}, with the important difference that 
now, both the black hole mass and its angular momentum are allowed
to increase. This leads to additional metric coefficients -- the $\phi$ 
component of the shift vector -- to be non zero, making the whole numerical 
treatment appreciably more difficult than in the case of the Schwarzschild 
metric. The spacetime metric is hence approximated at each time step
by a stationary (Kerr) black hole metric of varying mass $M_{\mathrm{BH}}$ 
{\it and} angular momentum $J_\mathrm{BH}$. We note that both quantities
increase very slowly during the evolution. Details of our procedure as 
well as code tests for rotating black holes will be reported in \citet{daigne:02}.

\section{Results}

Each of our four initial models is evolved three times. In the first series of
runs $M_{\mathrm{BH}}$ and $J_\mathrm{BH}$ are kept constant, hence the spacetime 
is held fixed. In the second series of runs only $M_{\mathrm{BH}}$ is allowed 
to increase while in the third series of runs both, $M_{\mathrm{BH}}$ and 
$J_\mathrm{BH}$ are allowed to increase. The growth rate is monitored through 
the mass and angular momentum transfer from the disk to the black hole across 
the innermost grid zone. Since in realistic disks angular momentum is transported 
outwards by dissipative processes or removed from the system by gravitational radiation, 
we adopt a conservative value for the efficiency of the angular momentum transfer, 
assuming that only 20\% of the angular momentum of the accreted material is 
transferred to the black hole. Results for other values of the efficiency are 
reported in \citet{daigne:02}.

The evolution of the mass flux $\dot{m}=2\pi\int_{0}^{\pi}\sqrt{-g}Dv^{r} d\theta$
for all twelve simulations is plotted in Fig.~1. In this expression $g$
is the determinant of the Kerr metric, $v^{r}$ is the radial component of the
3-velocity and $D=\rho\Gamma$ is the relativistic mass density, $\rho$ being the
rest-mass density and $\Gamma$ the Lorentz factor. At early times the mass flux 
evolution for all three series in each model is exactly the same, irrespective 
of the increase of the black hole mass and spin being taken into account or not.
For the first series of runs the mass flux reaches a stationary regime 
where $\dot{m}\propto\Delta W_\mathrm{in}^{4}$ \citep{font:02}, 
which is the theoretical expectation for $\gamma=4/3$ \citep{abramowicz:78}. 
Since we consider only models with fixed ratios $M_\mathrm{D}/M_\mathrm{BH}$ 
and $\Delta W_\mathrm{in}/W_\mathrm{cusp}$, the stationary mass flux is different 
for each model. However our results are not modified when all models have the same 
initial $\dot{m}$ (see \citet{daigne:02}).

For constant angular momentum disks (models A and C; $\alpha=0$) the time evolution 
of the system changes dramatically when the spacetime dynamics is taken into account, 
i.e.  when either $M_{\mathrm{BH}}$ only or $M_{\mathrm{BH}}$ {\it and} 
$J_\mathrm{BH}$ increase. In either case the runaway instability, reflected in the 
rapid growth of the mass accretion rate, appears on a dynamical timescale of 
$\sim 4\ t_{\mathrm{orb}} \sim 6.5$ ms for model A ($\sim 8\ t_{\mathrm{orb}} \sim 12$ 
ms for model C). The time derivative of the mass flux also increases, which implies 
the rapid divergence of $\dot{m}$. Comparing the case where only $M_{\mathrm{BH}}$ 
changes with the case where both $M_{\mathrm{BH}}$ and $J_\mathrm{BH}$ change, the 
instability appears slightly later in the latter case (dotted line in Fig.~1), the 
qualitative behaviour of the mass flux being, however, identical.

The main differences appear in the case of non-constant angular momentum disks
(models B and D). In Fig.~1 it becomes apparent that a slight increase 
of the specific angular momentum outwards (well below the Keplerian 
limit $\alpha=0.5$) strongly stabilizes the disk and completely suppresses the 
runaway instability. During the accretion process, the material at the inner edge 
of the disk is replaced by new material with a higher angular momentum. Therefore,
the centrifugal barrier which acts against accretion is much more efficient.
We note that the long-term behaviour of the mass flux evolution simply reflects 
the fact that the initial accretion rate is non zero (highly super-Eddington: 
$\sim 2.5\ \mathrm{M_{\odot}/s}$ for model B, $\sim 1.3\ \mathrm{M_{\odot}/s}$ 
for model D). Hence, by integrating $\dot{m}$ along the entire time of the simulation, 
it is possible to check that the total mass transferred becomes, asymptotically, 
of the order of the initial mass of the disk, which results in the end of the 
accretion process. This effect is more pronounced for smaller disk-to-hole mass 
ratios (model D). Notice that the mass flux decreases during the long-term 
evolution. This is due to the fact that for non-constant angular momentum 
disks (models B and D), the accretion process makes the cusp move faster towards 
the black hole than the inner radius of the disk (which is precisely why the 
instability is physically suppressed). Therefore, the potential at the inner 
edge $\Delta W_\mathrm{in}$ decreases with time and, consequently, also the 
mass flux. For this reason, accretion becomes increasingly slower, which makes 
numerically difficult to follow the evolution until the entire disk has been 
fully accreted. 

In Fig.~2 we plot the time evolution of the mass of the disk for all twelve
simulations. The unstable behaviour of models A and C, with $\alpha=0$, is 
characterized by the sudden decrease of $M_{\mathrm{D}}$ in less than $\sim 6-10$
orbital periods. By the end of the simulation the disk and black hole masses 
are $M_\mathrm{D}\sim 0.13$ M$_{\odot}$ and $M_{\mathrm{BH}}\sim 4.88$ 
M$_{\odot}$ for model A ($M_\mathrm{D}\sim 0.013$ M$_{\odot}$ and $M_{\mathrm{BH}}
\sim 2.74$ M$_{\odot}$ for model C), which means that 95\% of the disk mass has 
already been accreted. However, when the spacetime is held fixed (solid line), 
the mass of the disk for model A at $t\sim 200\ t_\mathrm{orb}$ is still 
$M_\mathrm{D}\sim 0.1$ M$_{\odot}$ (correspondingly, $M_\mathrm{D}\sim 0.015$ 
M$_{\odot}$ at $t\sim 100\ t_\mathrm{orb}$ for model C). This means that the 
accretion timescale is at least 30 and 10 times longer, for models A and C 
respectively, than the runaway (dynamical) timescale $t_\mathrm{run}$. On the 
other hand, for models B ($\alpha=0.1$) and D ($\alpha=0.075$), the stability 
properties of the non-constant distribution of angular momentum are dramatically 
reflected in the small loss of the mass of the disk throughout the evolution. 
At the end of the simulation the mass loss is only $\sim 2$\% for model B and 
$\sim 14$\% for model D (the lines are almost indistinguishable). The final masses 
of the black holes are $M_{\mathrm{BH}}\sim 2.55$ M$_{\odot}$ and $\sim 2.54$ 
M$_{\odot}$ for models B and D, respectively.

The transfer of mass and angular momentum from the disk to the black hole leads
to the gradual increase of the rotation law of the initially non-rotating black 
hole. In Fig.~3 we plot the time evolution of the black hole spin 
$a=J_\mathrm{BH}/M_{\mathrm{BH}}^2$ for both the stable and unstable cases. The stable 
disks, models B and D, only transfer a very small fraction of their angular momentum 
to the black hole.  At the end of the simulation the black hole spin is 
$a=0.014$ for model B and $a\sim 0.005$ for model D. On the contrary, for the 
unstable disks the transfer of angular momentum is considerably higher, accelerating 
rapidly during the runaway instability. Therefore, in $\sim 6-10$ orbital periods 
the initial Schwarzschild black hole turns into a mildly (slowly) rotating Kerr 
black hole with $a\sim 0.4$ ($a\sim 0.05$) for model A (model C).

\section{Discussion}

By means of time-dependent hydrodynamical simulations in general relativity we 
have shown that the runaway instability of (non self-gravitating) thick disks 
around (rotating) black holes \citep{abramowicz:83} is strongly suppressed when 
the distribution of angular momentum in the disk is non constant. A small 
increase outwards of the specific angular momentum distribution, well below
the Keplerian limit, could result in a strong stabilizing effect. In particular, 
we have demonstrated that for disks whose angular momentum follows a power-law, 
$l\propto r^{\alpha}$, the case $\alpha=0$ is unstable for both, $M_{\mathrm D}/
M_\mathrm{BH}=1$ and 0.1, while disks with small non-zero values of $\alpha$ are
stable ($\alpha=0.1$ for $M_{\mathrm D}/M_\mathrm{BH}=1$ and $\alpha=0.075$ for
$M_{\mathrm D}/M_\mathrm{BH}=0.1$). This result is in complete agreement with 
previous studies based on stationary sequences of equilibrium configurations, 
either using a pseudo-Newtonian potential \citep{daigne:97} or in general 
relativity \citep{abramowicz:98}. For the first time we have shown this effect 
with a time-dependent relativistic calculation, and we have been able to estimate 
the lifetime of a thick disk orbiting a black hole in both, the unstable and 
stable cases.

According to our results the runaway instability is most likely avoided in 
non rigidly rotating disks formed in the coalescence of a binary neutron star 
system or in the gravitational collapse of a massive star. If the growth of 
non-axisymmetric modes in the disk by the Papaloizou-Pringle instability is 
suppressed by the accretion process itself, as suggested by \citet{blaes:87}, 
systems consisting of a Kerr black hole surrounded by a high density torus may 
then be long lived. The lifetime is probably controlled  by the viscous timescale 
(a few seconds) rather than by the dynamical one, which may provide enough 
time for any plausible magneto-hydrodynamical process to efficiently transfer 
part of the energy reservoir of the system to a relativistic outflow. Therefore, 
the most favoured current GRB models (see \citet{woosley:01}) can indeed be based on 
such central engines. This is especially important in the context of the so-called 
internal shock model for the prompt gamma-ray emission \citep{rees:94}, where the 
observed lightcurve reflects the activity of the central engine, with no 
modification of the timescales other than the time dilation due to the redshift. 
In particular, the central engine has to survive for a duration at least comparable 
with the observed duration of the GRB.

We note that our study does not include the self-gravity of the disk, despite its
relevance when $M_{\mathrm{D}}/M_{\mathrm{BH}}\sim 1$.  Successful attempts
to long-term stable simulations of thick disks orbiting black holes by solving 
the coupled system of Einstein and general relativistic hydrodynamic equations 
seem, however, out of the scope of present day numerical relativity codes 
(see, e.g.  \cite{fontlr} and references there in). According to previous studies 
\citep{khanna:92,nishida:96a,masuda:98}, based on either time-dependent simulations
with pseudo-potentials or stationary models in general relativity, self-gravity 
has a strong destabilizing effect. However, this effect has to decrease for small 
disk-to-hole mass ratios. Therefore, we expect that for $M_\mathrm{D}/M_\mathrm{BH}
\la 0.1$ the stabilizing effect of non-constant angular momentum distribution 
demonstrated in this study could overcome the destabilizing effect of the much 
less relevant disk's self-gravity.

A comprehensive study of a broader sample of initial models accounting for
different disk-to-hole mass ratios and initial accretion rates will be presented 
elsewhere \citep{daigne:02}. We further plan to perform a careful study of the 
critical value of $\alpha$ separating stable and unstable disks and to find its 
dependence on the disk-to-hole mass ratio and on the black hole spin.

\acknowledgments
We are grateful to L.~Rezzolla and O.~Zanotti for interesting comments.
J.A.F. acknowledges financial support from a EU Marie Curie fellowship 
(HPMF-CT-2001-01172) and from the Spanish Ministerio de Ciencia y Tecnolog\'{\i}a 
(grant AYA 2001-3490-C02-01). F.D. acknowledges financial support from a 
postdoctoral fellowship from the French Spatial Agency (CNES).

\bibliographystyle{apj}

\begin{figure}
\begin{center}
\epsfxsize=6.9in
\leavevmode
\epsffile{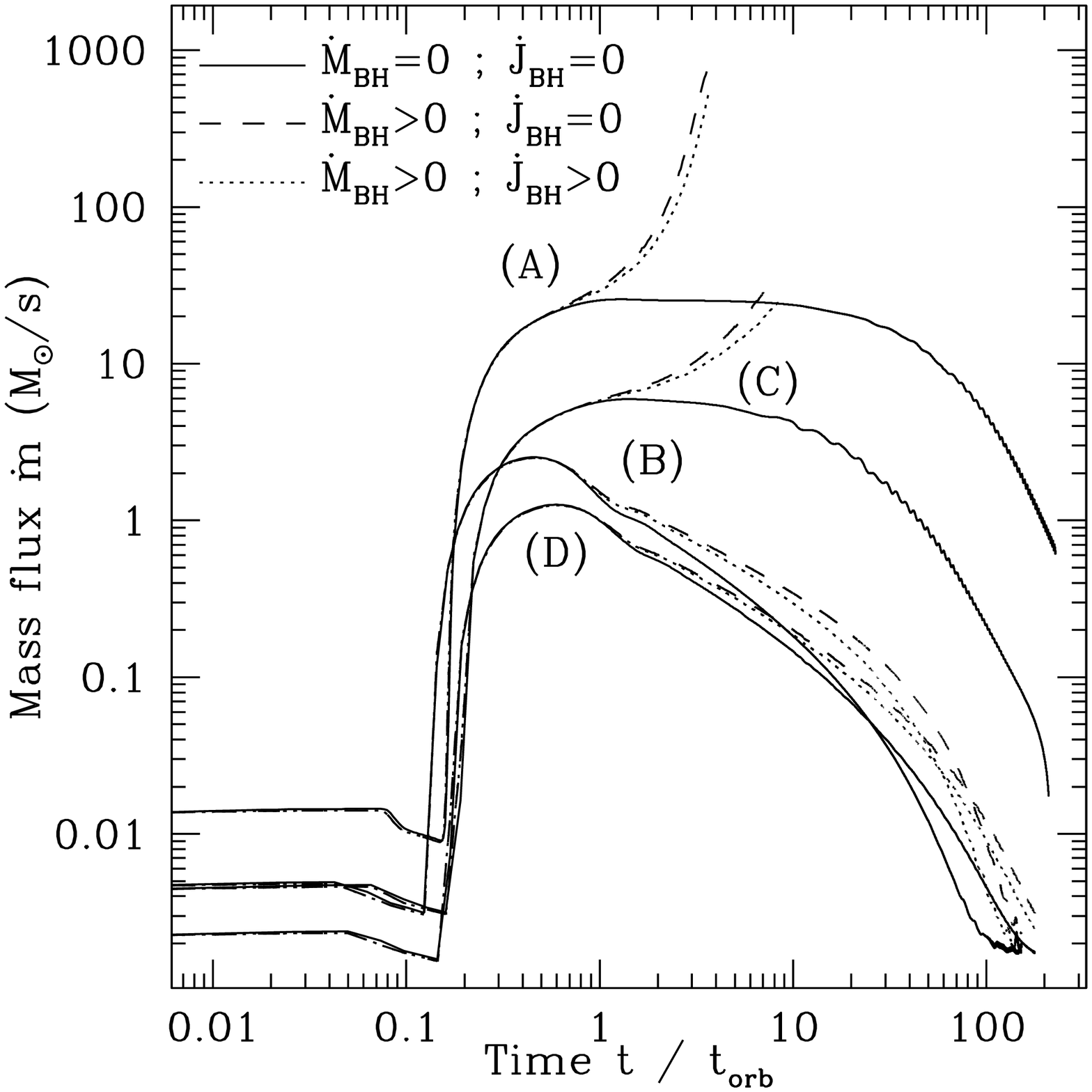}
\end{center}
\caption{
Time evolution of the mass flux. The solid lines correspond to
evolutions where the spacetime is held fixed, while the
dashed and dotted lines correspond to the cases where only the growth
of $M_{\mathrm{BH}}$ is allowed, and where both $M_{\mathrm{BH}}$
and $J_\mathrm{BH}$ increase, respectively.}
\label{fig1}
\end{figure}

\begin{figure}
\begin{center}
\epsfxsize=6.9in
\leavevmode
\epsffile{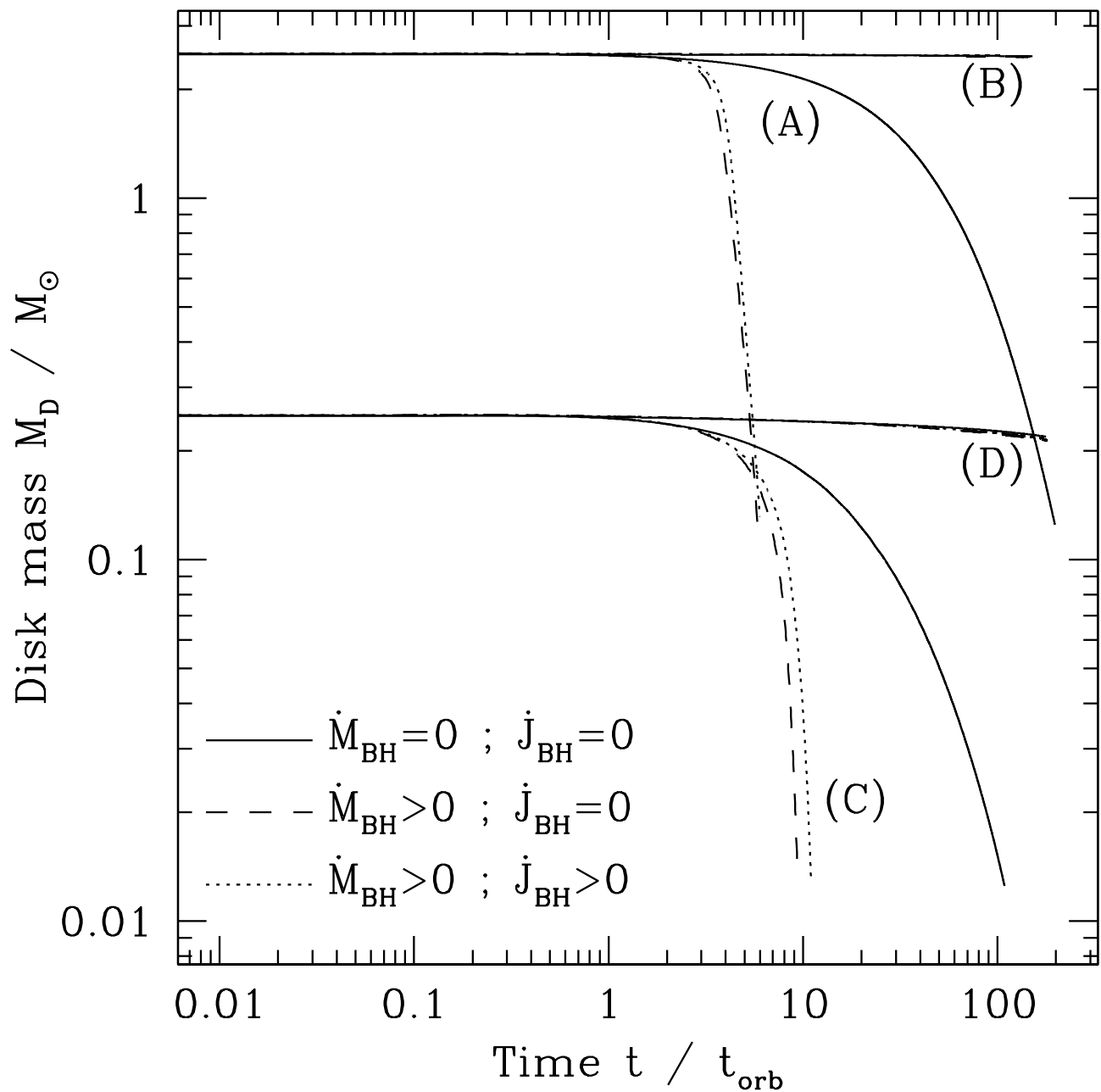}
\end{center}
\caption{
Time evolution of the mass of the disk. The line style follows the same
convention as in Fig.~1. The non-constant angular momentum disks, 
models B and D, remain stable throughout the whole simulation.} 
\label{fig2}
\end{figure}

\begin{figure}
\begin{center}
\epsfxsize=6.9in
\leavevmode
\epsffile{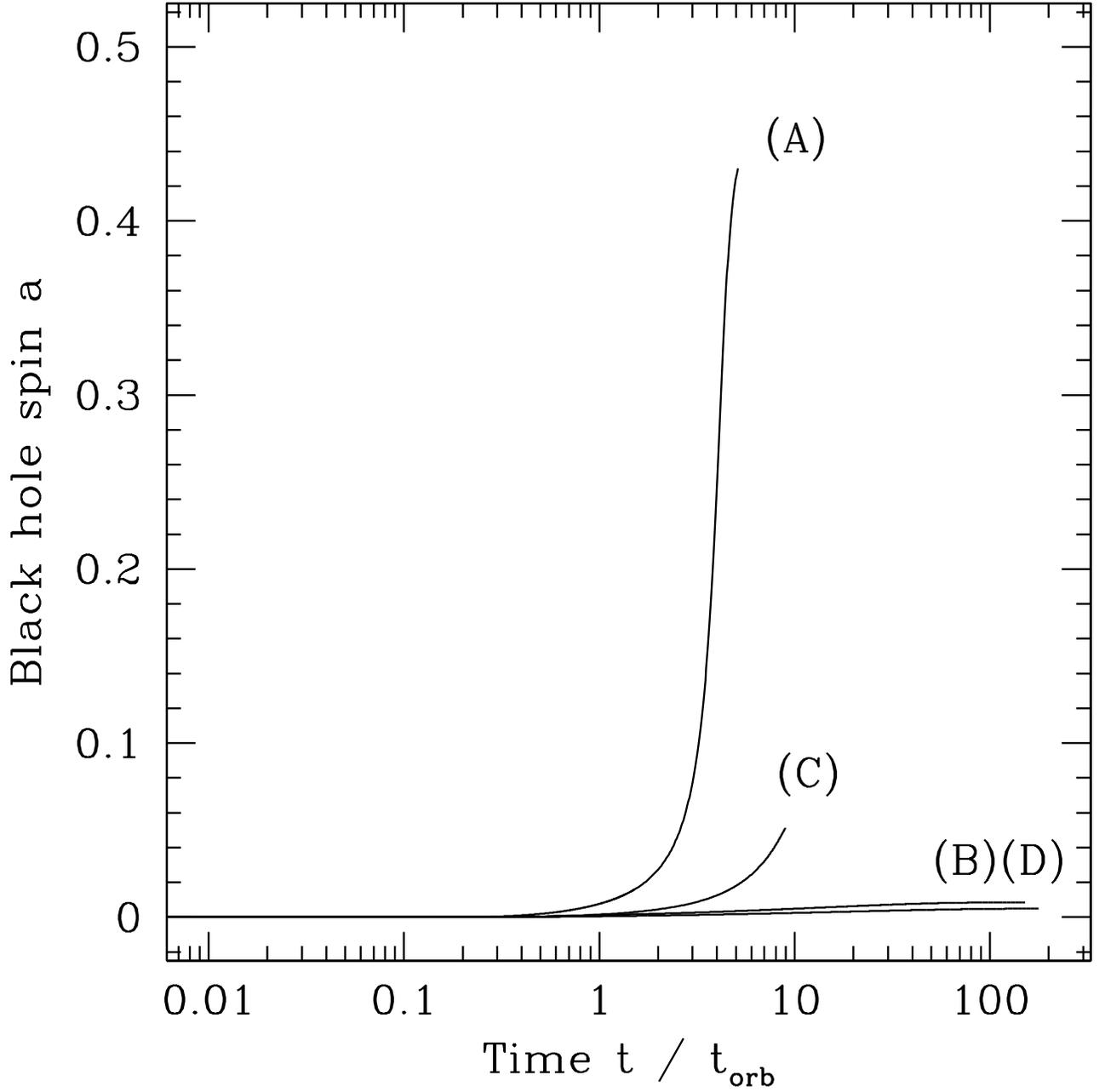}
\end{center}
\caption{
Time evolution of the spin of the black hole for the simulations in which 
the transfer of mass and angular momentum from the disk is allowed.}
\label{fig3}
\end{figure}

\end{document}